\documentclass[]{aa}
\usepackage{graphicx}
\usepackage{natbib}

\usepackage{color}

\usepackage{amssymb}

\usepackage[varg]{txfonts}

\begin{document}

\definecolor{bleu}{rgb}{.3,0.2,1.0}
\definecolor{red}{rgb}{1.,0.2,0.2}

\title{Comparison of theoretical and observed Ca~{\sc ii}~8542 Stokes profiles in quiet regions at the centre of the solar disc}

\author{J. Jur\v{c}\'{a}k
        \inst{1}
        \and
        J. \v{S}t\v{e}p\'{a}n
        \inst{1}
        \and
        J. Trujillo Bueno
        \inst{2, 3, 4}
        \and
        M. Bianda
        \inst{5}}

\institute{Astronomical Institute of the Academy of Sciences, Fri\v{c}ova  298, 25165 Ond\v{r}ejov, Czech Republic
  \and
  Instituto de Astrof\'{\i}sica de Canarias, E-38205 La Laguna, Tenerife, Spain
  \and
  Departamento de Astrof\'{\i}sica, Facultad de F\'\i sica, Universidad de La Laguna, E-38206 La Laguna, Tenerife, Spain
  \and
  Consejo Superior de Investigaciones Cient\'{\i}ficas, Spain
  \and
  Istituto Ricerche Solari Locarno, via Patocchi, 6605 Locarno-Monti, Switzerland}

\date{Received ; accepted }

\abstract
{Interpreting the Stokes profiles observed in quiet regions of the solar chromosphere is a challenging task. The Stokes $Q$ and $U$ profiles are dominated by the scattering polarisation and the Hanle effect, and these processes can only be correctly quantified if 3D radiative transfer effects are taken into account. Forward-modelling of the intensity and polarisation of spectral lines using a 3D model atmosphere is a suitable approach in order to statistically compare the theoretical and observed line profiles.}
  { Our aim is to present novel observations of the Ca~{\sc ii}~8542~\AA\ line profiles in a quiet region at the centre of the solar disc and to quantitatively compare them with the theoretical Stokes profiles obtained by solving the problem of the generation and transfer of polarised radiation in a 3D model atmosphere. We aim at estimating the reliability of the §3D model atmosphere, excluding its known lack of dynamics and/or insufficient density, using not only the line intensity but the full vector of Stokes parameters.}
 {We used data obtained with the ZIMPOL instrument at the Istituto Ricerche Solari Locarno (IRSOL) and compared the observations with the theoretical profiles computed with the PORTA radiative transfer code, using as solar model atmosphere a 3D snapshot taken from a radiation-magnetohydrodynamics simulation. The synthetic profiles were degraded to match the instrument and observing conditions.}
{The degraded theoretical profiles of the Ca~{\sc ii}~8542 line are qualitatively similar to the observed ones. We confirm that there is a fundamental difference in the widths of all Stokes profiles: the observed lines are wider than the theoretical lines. We find that the amplitudes of the observed profiles are larger than those of the theoretical ones, which suggests that the symmetry breaking effects in the solar chromosphere are stronger than in the model atmosphere. This means that the isosurfaces of temperature, velocity, and magnetic field strength and orientation are more corrugated in the solar chromosphere than in the currently available 3D radiation-magnetohydrodynamics simulation.}
{}


\maketitle

%
%

\section{Introduction}
\label{introduction}

The Ca~{\sc ii}~8542\,\AA\ line is one of the most promising lines in the magnetic diagnostics of both the quiet and active solar chromosphere. It is difficult to interpret the polarised spectra of this line because it forms under non-local thermal equilibrium (NLTE) conditions and because its linear polarisation profiles are significantly affected by the breaking of the axial symmetry produced by horizontal atmospheric inhomogeneities, both in active and quiet regions \citep[see][for more details]{Stepan:2016}. While the longitudinal component of the magnetic field can be well estimated from the circular polarisation profile of the line \citep[e.g.][]{CruzRodriguez:2012}, the linear polarisation is affected by the joint action of the Zeeman, Hanle, and symmetry breaking effects in active regions and by the Hanle and symmetry breaking mechanisms in quiet regions. Only in active regions are the ensuing Stokes profiles dominated by the Zeeman effect and can be interpreted by means of NLTE inversion codes \citep[e.g.][]{socas:2000}.

The amplitude of the quiet-Sun linear polarisation signals in the disc centre predicted by radiative transfer calculations in 1D semi-empirical models are of the order of $10^{-4}$ to $10^{-3}$ (i.e. very weak), and the Zeeman signals can be expected to be of the same order of magnitude \citep[][]{manso:2010}. However, full 3D calculations show that we can expect $Q/I$ and $U/I$ scattering polarisation amplitudes of the order of up to $10^{-2}$ if the spatial resolution is sufficiently high \citep{Stepan:2016}.

In this paper, we show the first quantitative comparison of the theoretical calculations of \citet{Stepan:2016} and spectropolarimetric observations at the centre of the solar disc. The goal of this initial study is to find basic similarities and differences between the theoretical and observed Stokes profiles and to point out the possible reasons for the discrepancies.

\begin{figure*}[!t]
 \centering \includegraphics[width=0.95\linewidth]{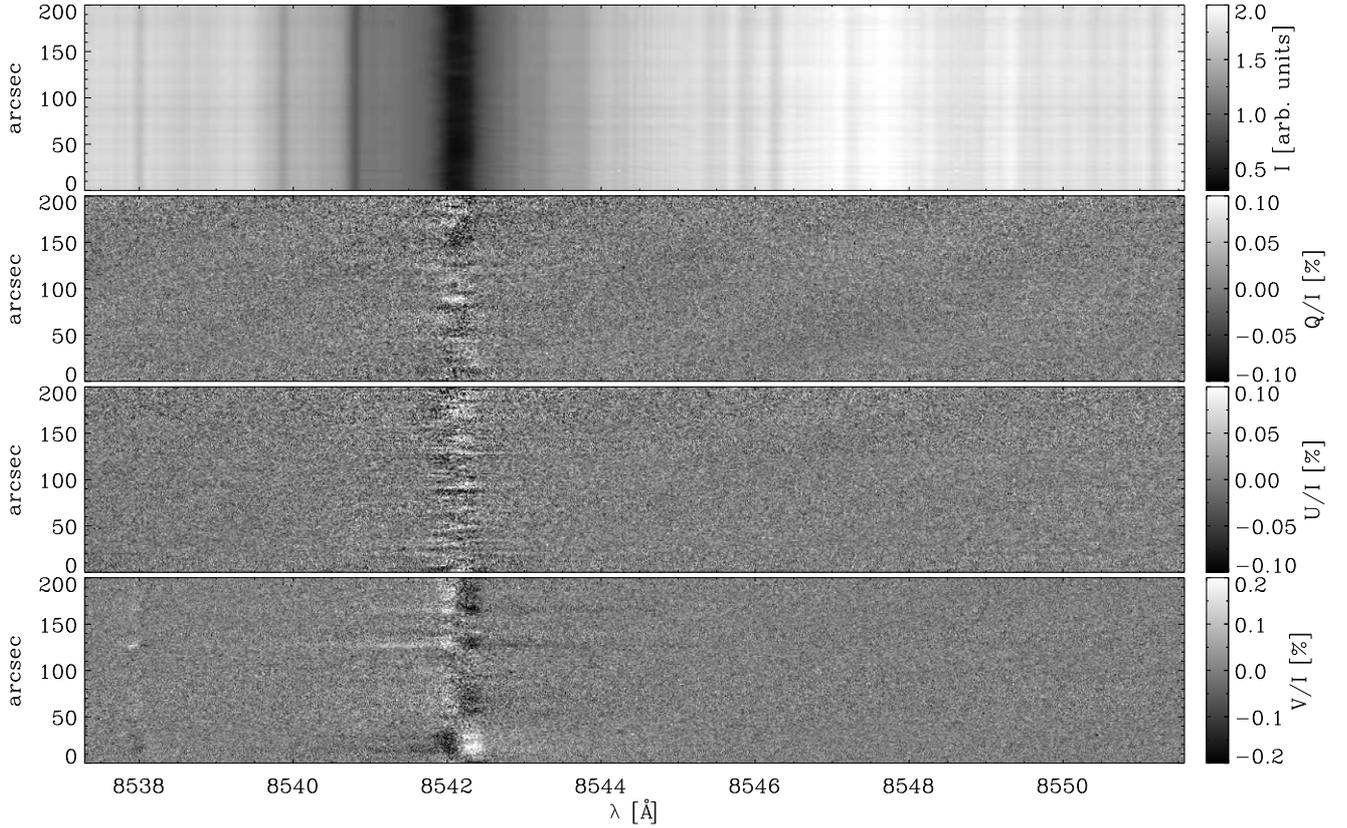}
 \caption{From top to bottom we show the Stokes $I$, $Q/I$, $U/I$, and $V/I$ profiles of the Ca~{\sc ii}~8542~\AA\ line along the spectrograph slit, observed in a quiet region at the centre of the solar disc using ZIMPOL at IRSOL.}
 \label{fig_obs}
\end{figure*}
\begin{figure}[!t]
 \centering \includegraphics[width=0.95\linewidth]{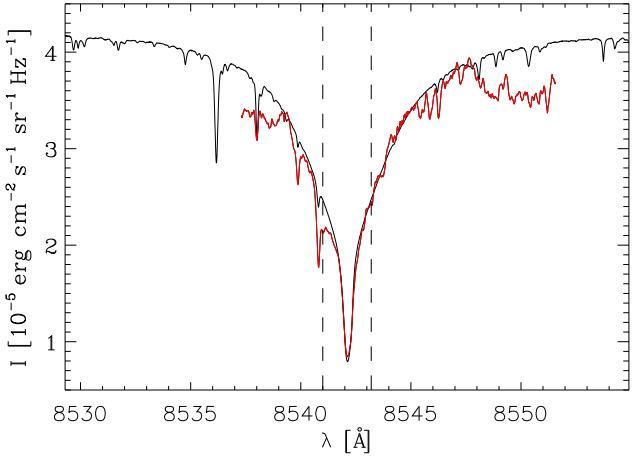}
 \caption{Comparison of the FTS atlas (black line) with the average Stokes~$I$ profile of the analysed observations (red line). The vertical dashed lines mark the wavelength region selected for our analysis.}
 \label{atlas_cal}
\end{figure}
\begin{figure}[!t]
 \centering \includegraphics[width=0.95\linewidth]{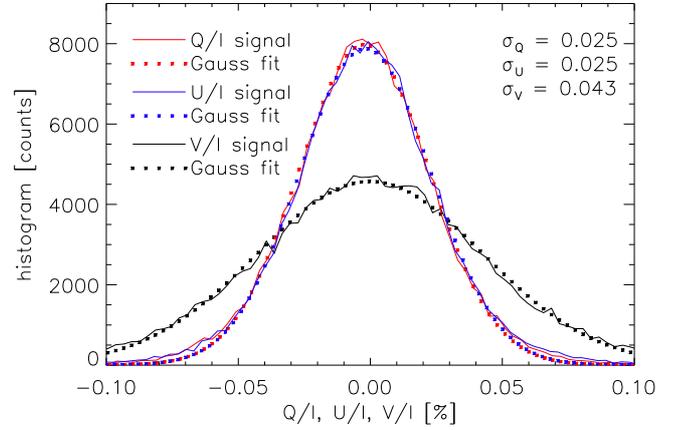}
 \caption{Distribution of noise of the fractional polarisation signals $Q/I$, $U/I$, and $V/I$ in the observations.}
 \label{noise_cal}
\end{figure}

\section{Observed profiles}
\label{obs}

Stokes profiles of the Ca~{\sc ii}~8542 \AA\ line were observed at the Istituto Ricerche Solari Locarno (IRSOL) using the ZIMPOL instrument \citep{ramelli:2010} attached to the Gregory-Coud\'{e} Telescope (aperture of 45~cm). The observations of the solar disc centre were taken at 15:23~UT on September 2, 2015. We took 100 exposures, each of one second. The data were corrected for flat-field, dark current, and they were filtered using a fast Fourier filter. Figure~\ref{fig_obs} shows the Stokes profiles of the Ca~{\sc ii}~8542 line, including the full wavelength range of the observations. There are 140 pixels along the spectrograph slit with a spatial sampling of 1.43\arcsec\ per pixel, while the spectral sampling of the observations is 13.7~m\AA. 

There are no slit-jaw images for these observations, and we do not know the exact location of the slit. There was no enhanced activity around the disc centre on September 2, 2015, and we did not identify any locally enhanced intensity signals that might be attributed to the solar network that the spectrograph slit must cross given its length of 200\arcsec\ and the typical size of supergranulation of 40\arcsec.

In order to calibrate the absolute wavelengths and intensities of the observations, we compared the averaged Stokes $I$ profile with the FTS atlas intensity profile \citep{Neckel:1999}. This is shown in Fig.~\ref{atlas_cal}. The telluric lines are more prominent in the IRSOL observations because of the lower altitude and the greater humidity in the air. Far away from the Ca~{\sc ii}~8542 line core, the decrease of the average observed profile is caused by instrumental effects. This does not influence our analysis, as we focus only on the line core depicted by the dashed lines.

We used the polarimetric data out of the line core, where the polarimetric signals are dominated by noise (see Fig.~\ref{fig_obs}), to estimate the noise level in the $Q/I$, $U/I$, and $V/I$ profiles. In Fig.~\ref{noise_cal} we show that the noise distribution is well fitted by Gaussian functions; the $\sigma$ of the fits is comparable for the linear polarisation signals ($\sigma_{Q,U}=0.025 \%$) and the circular polarisation signals have $\sigma_{V}=0.043~\%$. The Gaussian distributions with these standard deviations were used to create the artificial noise needed to degrade the synthetic Stokes profiles of the Ca~{\sc ii}~8542~\AA\ line, so as to make a proper comparison with the observed profiles.

\begin{figure*}[!t]
 \sidecaption
 \includegraphics[width=0.72\linewidth]{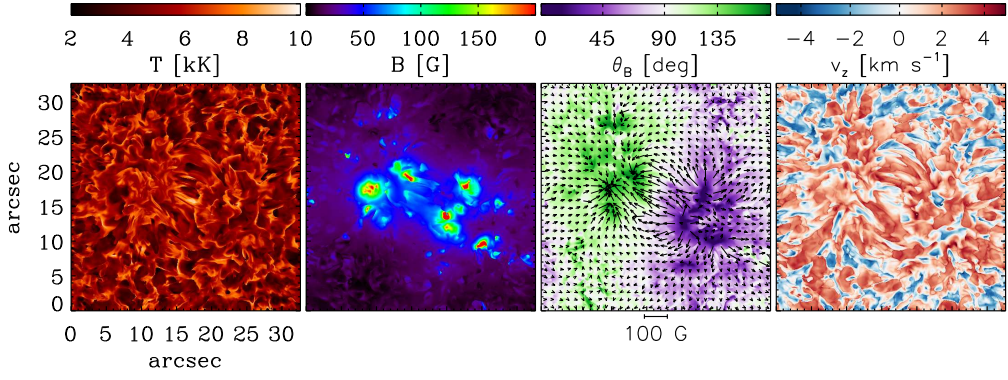}
 \caption{From left to right, we show  maps of temperature, magnetic field strength, inclination, and vertical velocity (with $v_z>0$ corresponding to downflow) in the MHD simulation snapshot, at the heights of the line-centre optical depth unity in the Ca~{\sc ii}~8542 line. The arrows in the inclination map indicate the magnetic field azimuth, and their length is proportional to the horizontal component of the magnetic field vector.}
 \label{fig_mhd}
\end{figure*}
\begin{figure*}[!t]
 \sidecaption
 \includegraphics[width=0.72\linewidth]{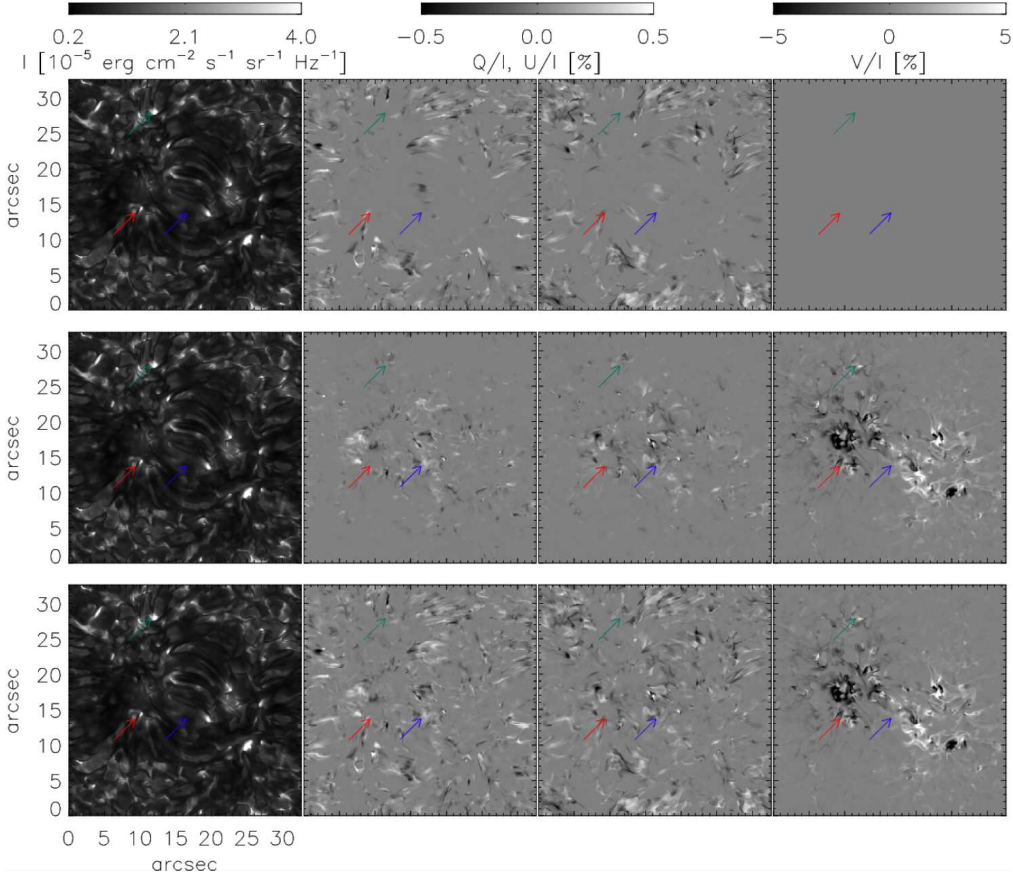} 
 \caption{From left to right, we show maps of the line centre ($\lambda=8542.13$~\AA) Stokes $I$, $Q/I$, and $U/I$ signals. The $V/I$ signals are shown at the near-blue wing ($\lambda = 8542$~\AA) of the line. We show the polarisation signals resulting from the scattering polarisation and the Hanle effect (top row), the Zeeman effect (middle row), and the joint action of the scattering polarisation and the Hanle and Zeeman effects (bottom row). The arrows point to the pixels with representative Stokes profiles, shown in Fig.~\ref{porta_profiles}.}
 \label{fig_porta}
\end{figure*}

\begin{figure*}[!t]
 \centering \includegraphics[width=0.95\linewidth]{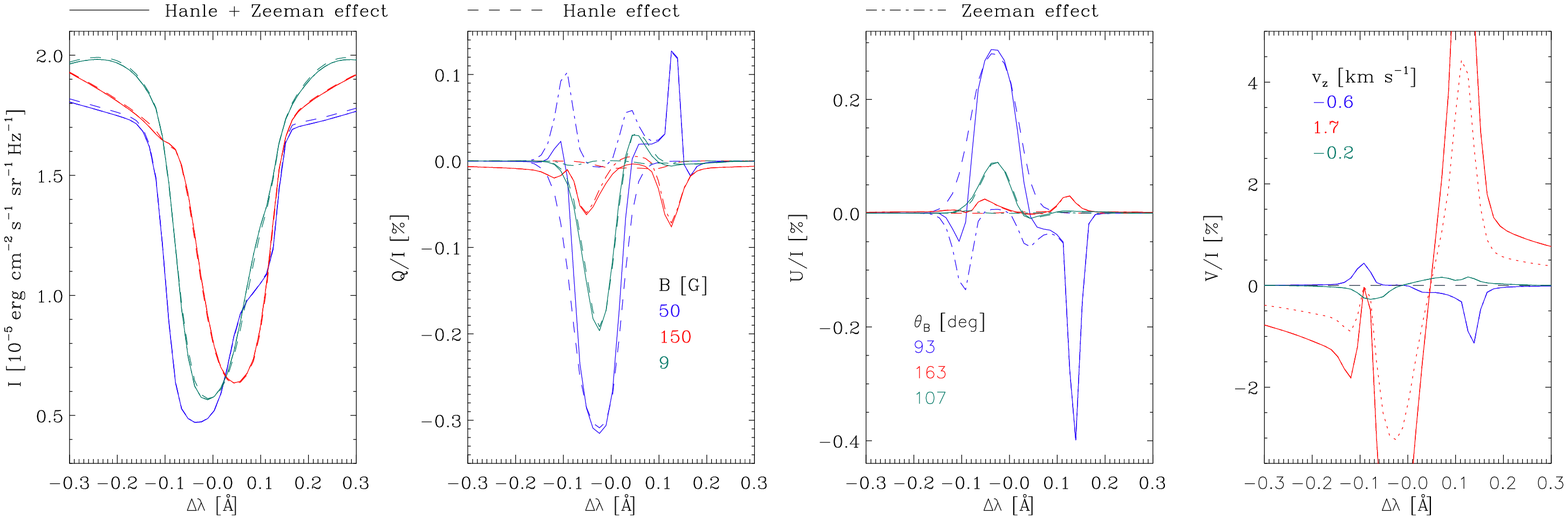}
 \caption{Stokes profiles of the Ca~{\sc ii}~8542 line at the pixels marked by the coloured arrows in Fig.~\ref{fig_porta}. The line styles are explained above the plots, and the plasma parameters at individual pixels are depicted in the plots.}
 \label{porta_profiles}
\end{figure*}

\begin{figure}[!b]
 \centering \includegraphics[width=1\linewidth]{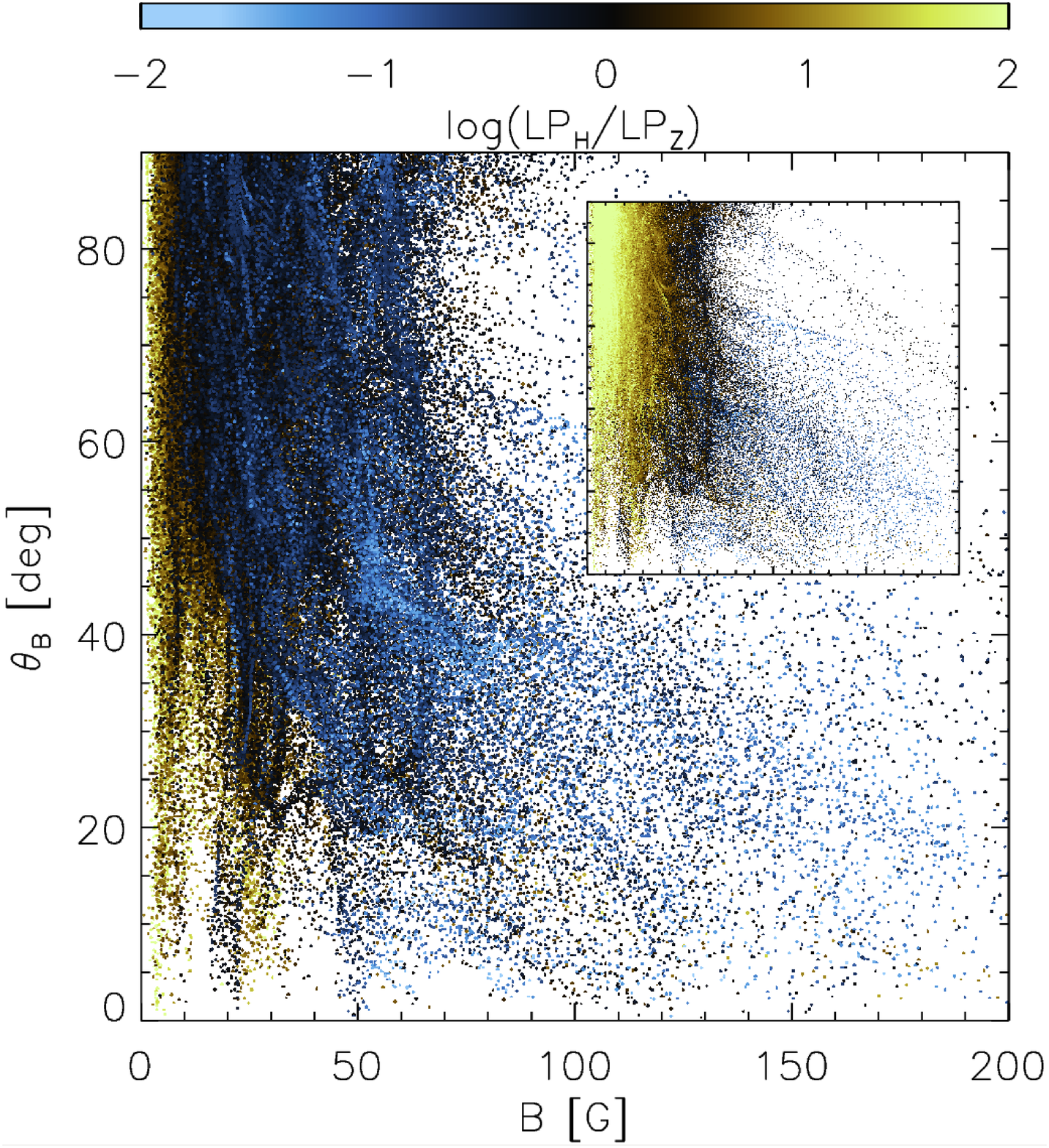}
 \caption{Scatter plot of the ratio of the maximum linear polarisation signal produced by scattering polarisation and the Hanle effect (LP$_\mathrm{H}$) and the Zeeman effect (LP$_\mathrm{Z}$) as a function of the magnetic field strength and inclination. In the main plot, the preference is given to points with lower (less common) values of $\mathrm{LP_H/LP_Z}$. In the inset, the preference is given to the pixels with higher values of $\mathrm{LP_H/LP_Z}$. The scale of the inset axes is identical to the main plot.}
 \label{lp_ratio}
\end{figure}

\section{Theoretical profiles}

The theoretical profiles depend on the model atmosphere, which is described in Sect.~\ref{model_atm}. The synthesis of the profiles was reported by \citet{Stepan:2016}, who applied the radiative transfer code PORTA \citep{Stepan:2013}, as described in Sect.~\ref{profile_synth}. To compare the observed and theoretical profiles of the Ca~{\sc ii}~8542 \AA\ line, we have to degrade the spatial and spectral resolutions of the synthetic Stokes profiles to match the observing conditions at IRSOL. This step is described in Sect.~\ref{degradation}.

\subsection{Model atmosphere}
\label{model_atm}

We used the publicly available 3D snapshot model resulting from the MHD simulations by \citet{Carlsson:2016}, carried out with the Bifrost code \citep{Gudiksen:2011}. The computational box has $24 \times 24$~Mm$^2$ in the horizontal directions (equivalent to $32\farcs6 \times 32\farcs6$ on the centre of the solar disc), and it extends vertically from 2.4~Mm below the visible surface to the corona. The simulation box contains two patches of opposite polarities, with photospheric magnetic fields reaching 1.9~kG at the $\tau_{500}=1$ level and with magnetic field loops reaching chromospheric and coronal heights. This is described by \citet{Carlsson:2016} as an enhanced network region. In Fig.~\ref{fig_mhd} we show maps of plasma parameters at the heights of line-centre optical depth unity in the Ca~{\sc ii}~8542 line for this particular snapshot. 

Our study is based on snapshot 385 of the simulation run. For a more rigorous comparison with the observations, which have an exposure time of 100~s, the theoretical profiles would need to be averaged using ten consecutive snapshots, since the time interval between them is 10~s. Given the significant computing time such a calculation would require, we did not perform this averaging, and our theoretical polarisation amplitudes should therefore be understood as being the upper limit of the theoretical signals. Nevertheless, as we show below, the observed fractional polarisation amplitudes are still larger than the theoretical ones.

\begin{figure*}[!t]
 \centering \includegraphics[width=0.95\linewidth]{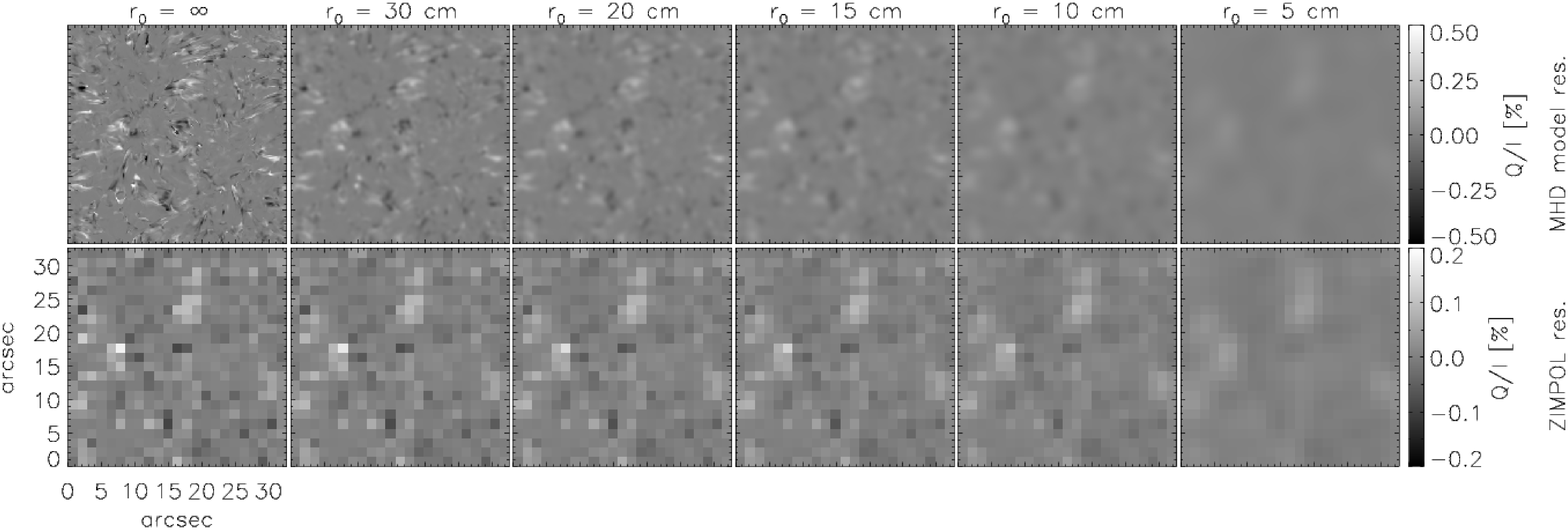}
 \caption{Synthetic $Q/I$ signals at the near-blue wing ($\lambda = 8542$~\AA) of the Ca~{\sc ii} line. The signals in the upper and lower rows correspond to the spatial resolution of the MHD simulation and of the ZIMPOL measurements, respectively. From left to right we show the influence of the atmospheric turbulence using different $r_0$ parameters.}
 \label{fig_deg}
\end{figure*}

\subsection{Profile synthesis}
\label{profile_synth}

The synthetic Stokes profiles used in this work were calculated by \citet{Stepan:2016} by applying the radiative transfer code PORTA \citep{Stepan:2013}, which solves the full 3D NLTE problem of the generation and transfer of scattering polarisation taking into account the symmetry-breaking effects caused by the thermal, dynamic, and magnetic structure of the 3D atmospheric model. The synthesis was carried out assuming complete frequency redistribution. We use the five-level Ca\,{\sc ii} model atom described in \citet{manso:2010} and also the population transfer collisional rates for all the transitions and the alignment transfer collisional rates for the allowed transitions specified in that paper.

The PORTA code allows us to compute the polarisation signal induced by the scattering polarisation and Hanle effect, the Zeeman effect, and the joint action of the scattering polarisation and the Hanle and Zeeman effects. In Fig.~\ref{fig_porta} we show the synthetic Stokes $I$, $Q/I$, $U/I$ and $V/I$ signals of the Ca~{\sc ii}~8542 \AA\ line for these alternatives. 

It is evident that scattering polarisation processes produce the strongest linear polarisation signals in the quietest regions of the simulation box, but the action of the Hanle effect diminishes the $Q$ and $U$ signals at the footpoints of the magnetic loop, that is to say, in regions where the magnetic field is strong and oriented along the line of sight (LOS). In this case, the contribution due to the transverse
Zeeman effect vanishes as well. In regions around the magnetic footpoints, where the magnetic field is inclined with respect to the LOS, we can
find strong linear polarisation signals due to the transverse Zeeman effect, while the scattering polarisation remains eliminated due to the
Hanle effect. The Zeeman effect is solely responsible for the circular polarisation signals. In the bottom row of Fig.~\ref{fig_porta}, we show the polarisation signals due to the joint action of the Zeeman effect, scattering polarisation, and the Hanle effect.

In Fig.~\ref{porta_profiles} we show examples of the characteristic Stokes profiles in regions dominated by the scattering polarisation (green lines) and the Zeeman effect (red lines). The blue lines show Stokes profiles in the region where scattering polarisation as well as the Hanle and Zeeman effects play a significant role. It is important to note that the Stokes profiles resulting from the joint action of scattering polarisation and the Hanle and Zeeman effects (solid lines) are not simply the sum of the scattering polarisation and the Hanle effect (dashed lines) and those caused by the Zeeman effect (dash-dotted lines). For a more detailed discussions of the role of the different polarising mechanisms, see \citet{Stepan:2016}

We also wish to stress that the simulation box contains magnetic fields stronger than 10~G at the formation height of the  Ca~{\sc ii}~8542 \AA\ line core (Hanle saturation regime for this line) in more than 82\% of the grid points. Despite this fact, the maximum linear polarisation signal caused by the scattering polarisation and the Hanle effect (LP$_\mathrm{H}$) is stronger than the maximum linear polarisation signal caused by the Zeeman effect (LP$_\mathrm{Z}$) in more than 85\% of the pixels. In Fig.~\ref{lp_ratio} we show the colour-coded ratio of LP$_\mathrm{H}$/LP$_\mathrm{Z}$ as a function of the magnetic field strength and inclination. We note that for given values of $B$ and $\theta_B$, there can be different values of LP$_\mathrm{H}$/LP$_\mathrm{Z}$ because of the different local gradients of the plasma parameters, and this cannot be well represented by the 2D scatter plot.

\subsection{Degradation}
\label{degradation}

In order to compare the observed profiles with the theoretical profiles presented in Sect.~\ref{profile_synth}, we need to take into account the degradation produced by the observing conditions and the instrumentation that was used. The first step is to take into account the seeing conditions during the observations. They can be quantified by the Fried parameter $r_0$, which can be interpreted as the diameter of a diffraction-limited telescope located outside of the Earth's atmosphere. Therefore, we used the Airy function for $\lambda=8542$~\AA{} and a given value of $r_0$ for the spatial point spread function (PSF) of the Earth's atmosphere. As the actual seeing conditions during the observations are unknown, we computed a set of PSFs (from $r_0=\infty$ - no degradation  to $r_0=5$~cm - poor seeing conditions).

The next step was to take the spatial PSF of the telescope into account, that is, an Airy function for $\lambda=8542$~\AA{} and $r=45$~cm. We omitted this step because the instrumental PSF was significantly narrower than the pixel size of the ZIMPOL camera and also narrower than any of the seeing-induced PSFs. To match the spatial sampling of the ZIMPOL camera, we rescaled the synthetic images from their original spatial resolution of $\sim$0.06\arcsec\ to 1.43\arcsec\ that matches the pixel size along the spectrograph slit. In Fig.~\ref{fig_deg} we show an example of how these steps of data degradation influence the $Q/I$ signals.

We also matched the wavelength sampling and the spectral resolution of the observations. To do this, we interpolated the synthetic profiles with a spline function with a sampling of 1 m\AA. Then, we convolved the synthetic profiles with the spectral PSF of the ZIMPOL instrument (at 8542~\AA, the spectral resolution is 15~m\AA) and re-sampled the synthetic profiles at the observed wavelengths. The last step of the degradation process was the addition of Gaussian white noise using the $\sigma$ values determined from observations (see Fig.~\ref{noise_cal}). 

Given that the 3D model contains an enhanced network, which does not resemble the observed quiet-Sun region, we selected only pixels where the synthetic $V/I$ signal (degraded with a PSF$_{r_0=10~\mathrm{cm}}$) was lower than 0.3\%. This threshold is related to the maximum observed $V/I$ signals.  In the following sections, we discuss only pixels that fulfil this criterion.

\begin{figure*}[!t]
 \centering \includegraphics[width=0.95\linewidth]{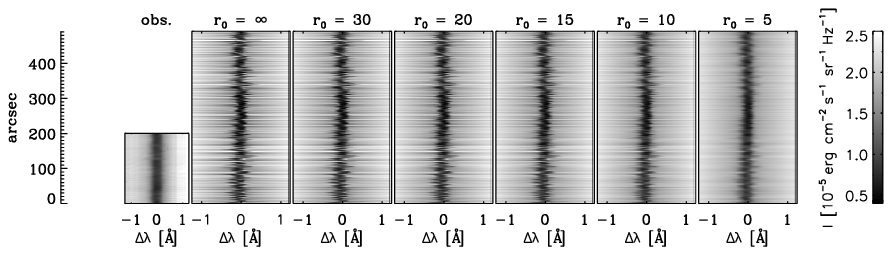}
 \centering \includegraphics[width=0.95\linewidth]{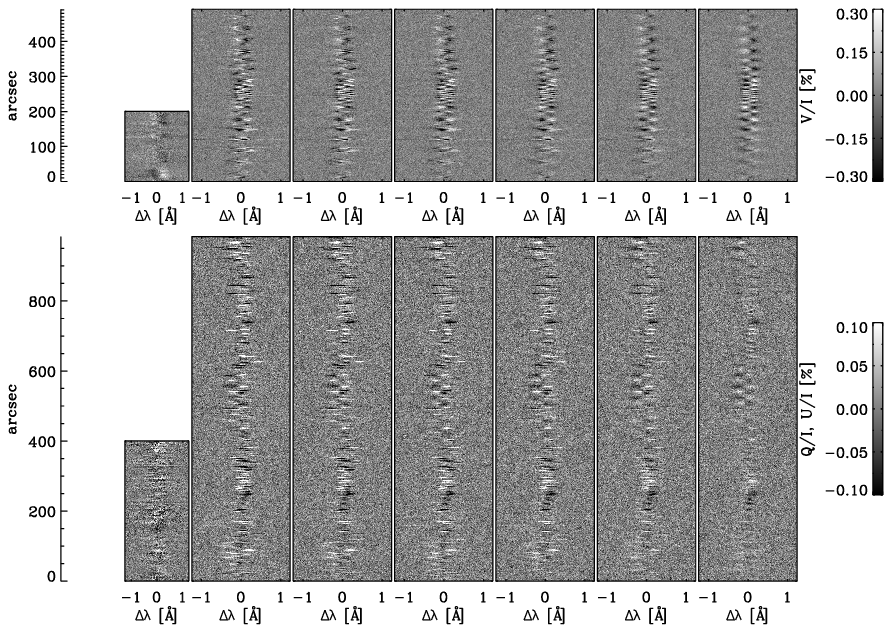}
 \caption{Comparison of the observed (left column) and synthetic (remaining columns) Stokes $I$ profiles (top row), Stokes $V/I$ profiles (middle row), and the $Q/I$ and $U/I$ profiles (bottom row). The synthetic profiles are more strongly spatially degraded the farther to the right they lie.}
 \label{fig_comp}
\end{figure*}

\section{Results}
\label{comparison}

In Fig.~\ref{fig_comp} we show the observed and theoretical profiles. The theoretical profiles are presented so as to resemble an observation made with a slit-based spectropolarimeter. Therefore, we can see a repetitive pattern. At the centre of the solar disc, the statistical properties of the $Q/I$ and $U/I$ profiles are the same. We therefore treated them together to increase the significance of the statistical sample of the linear polarisation profiles. The lower intensity of the blue wing of the observed Stokes $I$ profile of the Ca~{\sc ii}~8542~\AA\ line is caused by the atmospheric line blend (see Fig.~\ref{atlas_cal}) and cannot be reproduced in the synthetic profiles.

Visually, the synthetic profiles degraded with PSF$_{r_0=5~\mathrm{cm}}$ and PSF$_{r_0=10~\mathrm{cm}}$ match the observed signals best. These seeing conditions can be expected at the IRSOL observing site. The time averaging, which could not be taken into account in this work, would also further smoothen out the fine structures along the slit, as emphasized by \citet{Carlin:2017}. We note that the span of the  wavelength axis for the observed profiles is shortened by a factor of 0.65 to match the width of the theoretical profiles. This factor was pointed out by \citet{Stepan:2016}, who compared the theoretical profiles of the Ca~{\sc ii}~8542~\AA\ line with the FTS atlas. The differences in the line widths are discussed in Sect.~\ref{line_width}, the amplitudes of the observed and synthetic profiles are discussed in Sect.~\ref{line_amplitude}, and the shape of the profiles is described in Sect.~\ref{line_shape}

\subsection{Line width}
\label{line_width}

The discrepancy in the widths of the observed profiles and those resulting from the model atmosphere we used is well described for Stokes $I$ profiles for several spectral lines \citep[see e.g.][]{Pontieu:2015, Carlsson:2015, Stepan:2016}.

That the theoretical Stokes profiles of the Ca~{\sc ii}~8542~\AA\ line are narrower than the observed profiles can be interpreted as a lack of dynamics in the model atmosphere \citep[see also][for an alternative explanation]{Carlsson:2015}. We need to point out that the radiative transfer calculations were carried out without ad hoc parameters that cause line broadening, that is to say, neither micro- nor macroturbulent velocities were considered for the line synthesis. 

\begin{figure}[!t]
 \centering \includegraphics[width=0.95\linewidth]{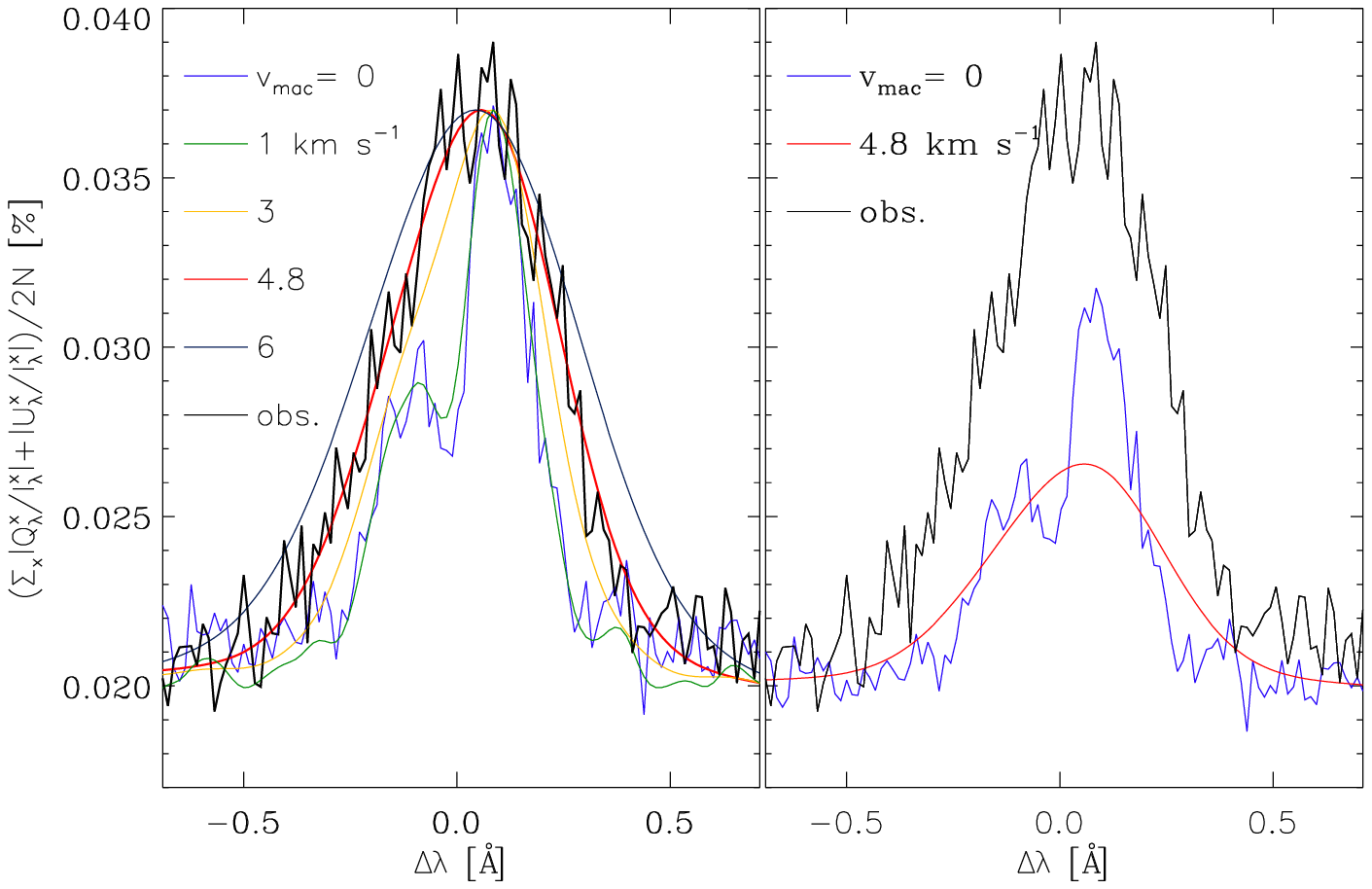}
 \caption{Comparison of the averaged $|Q/I|$ and $|U/I|$ profiles. Thick black curves correspond to the observations, coloured curves to the theoretical profile assuming PSF$_{r_0=10~\mathrm{cm}}$ convolved with the Gaussian functions representing macroturbulent velocities of different amplitudes. In the left panel, the maximum amplitudes are scaled to the observations. In the right panel, we show the actual amplitudes of the theoretical profiles.}
 \label{fig_macr}
\end{figure}

To quantify the missing dynamics, we compared the average profiles of the observed and synthetic linear polarisation profiles (we averaged $|Q/I|$ and $|U/I|$ profiles to prevent the cancellation of the signal). In Fig.~\ref{fig_macr} we show the average observed (black) and synthetic (blue) profiles. We chose a synthetic profile degraded with PSF$_{r_0=10~\mathrm{cm}}$. The effect of macroturbulence ($v_\mathrm{mac}$) is equivalent to the convolution of the theoretical profile with a Gaussian function whose width is proportional to the amplitude of $v_\mathrm{mac}$. 

In the left panel of Fig.~\ref{fig_macr}, we compare the theoretical profile broadened with different $v_\mathrm{mac}$ values with the observed profile, and we normalise the amplitudes of these convolved theoretical profiles to the amplitude of the observed profile. We find that the theoretical profile convolved with a Gaussian function with width equivalent to $v_\mathrm{mac}=4.8$~km~s$^{-1}$ provides the best match to the observed profile. In the right panel of Fig.~\ref{fig_macr}, we show the actual amplitudes of the averaged observed profile, the theoretical profile, and the theoretical profile broadened by $v_\mathrm{mac}$ of 4.8~km~s$^{-1}$.

Even though we did not make any numerical experiments with microturbulent velocities, we note that the value of $v_\mathrm{mac}$ we find is comparable to the values of $v_\mathrm{mic}$ that are commonly used when modelling chromospheric lines. The semi-empirical 1D models of the solar atmosphere have tabulated values of $v_\mathrm{mic}$ that are around $4-5$~km~s$^{-1}$ for the VAL models \citep{Vernazza:1981} and $3.5-13$~km~s$^{-1}$ for the FAL models \citep{Fontenla:1993} at an atmospheric height of 1500~km that approximately corresponds to the formation height of the Ca~{\sc ii}~8542 \AA\ line core. 

\begin{figure}[!t]
 \centering \includegraphics[width=0.95\linewidth]{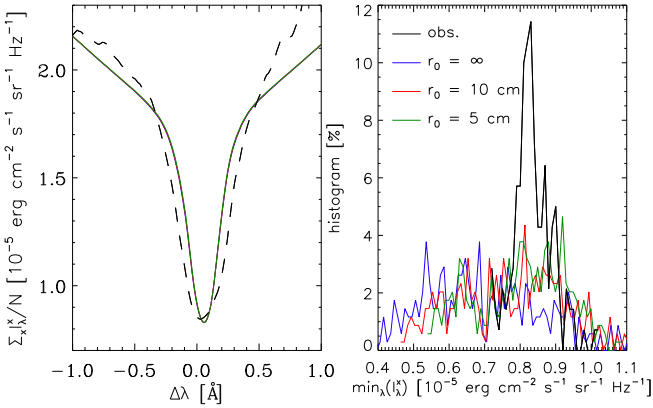}
 \caption{Left panel: Comparison of the shapes of the average observed (dashed line) and synthetic (solid line) intensity profiles. Right panel: Histograms of the maximum absolute intensities in the observed and synthetic profiles.}
 \label{fig_sig_int}
\end{figure}

\begin{figure}[!t]
 \centering \includegraphics[width=0.95\linewidth]{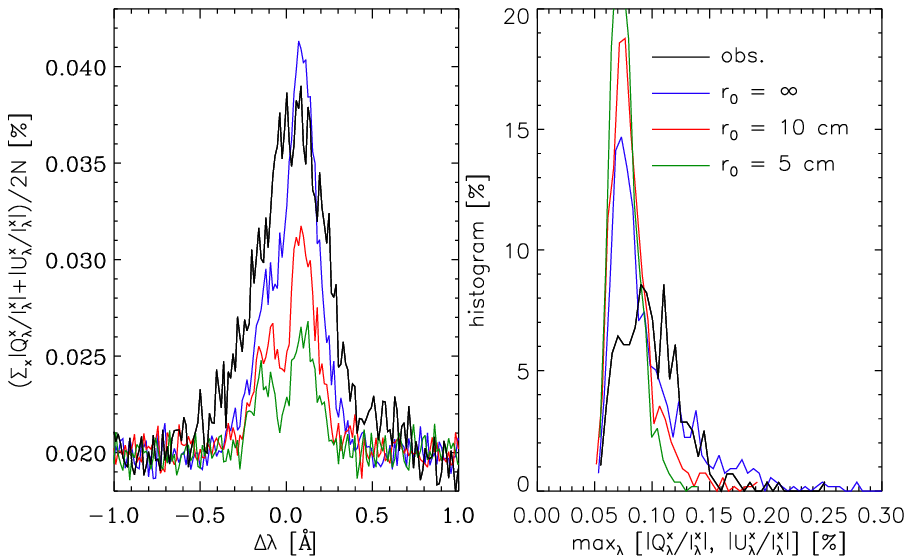}
 \caption{Left panel: Comparison of the shapes of the average absolute profiles corresponding to the observed and synthetic $Q/I$ and $U/I$ profiles. Right panel: Histograms of the absolute amplitudes of the $Q/I$ and $U/I$ profiles in the observed and synthetic data.}
 \label{fig_sig}
\end{figure}

\subsection{Profile amplitudes}
\label{line_amplitude}

In the left panel of Fig.~\ref{fig_sig_int}, we compare the average observed Stokes $I$ profile with the average synthetic profiles. As the Stokes $I$ profiles are positive everywhere, there are no cancellation effects, and the average synthetic profiles are independent of the spatial degradation. Except for the discrepancy in the width of the observed and synthetic profiles discussed in Sect.~\ref{line_width}, the achieved match in the line core intensity is very good.

\begin{figure*}[!t]
 \centering \includegraphics[width=0.95\linewidth]{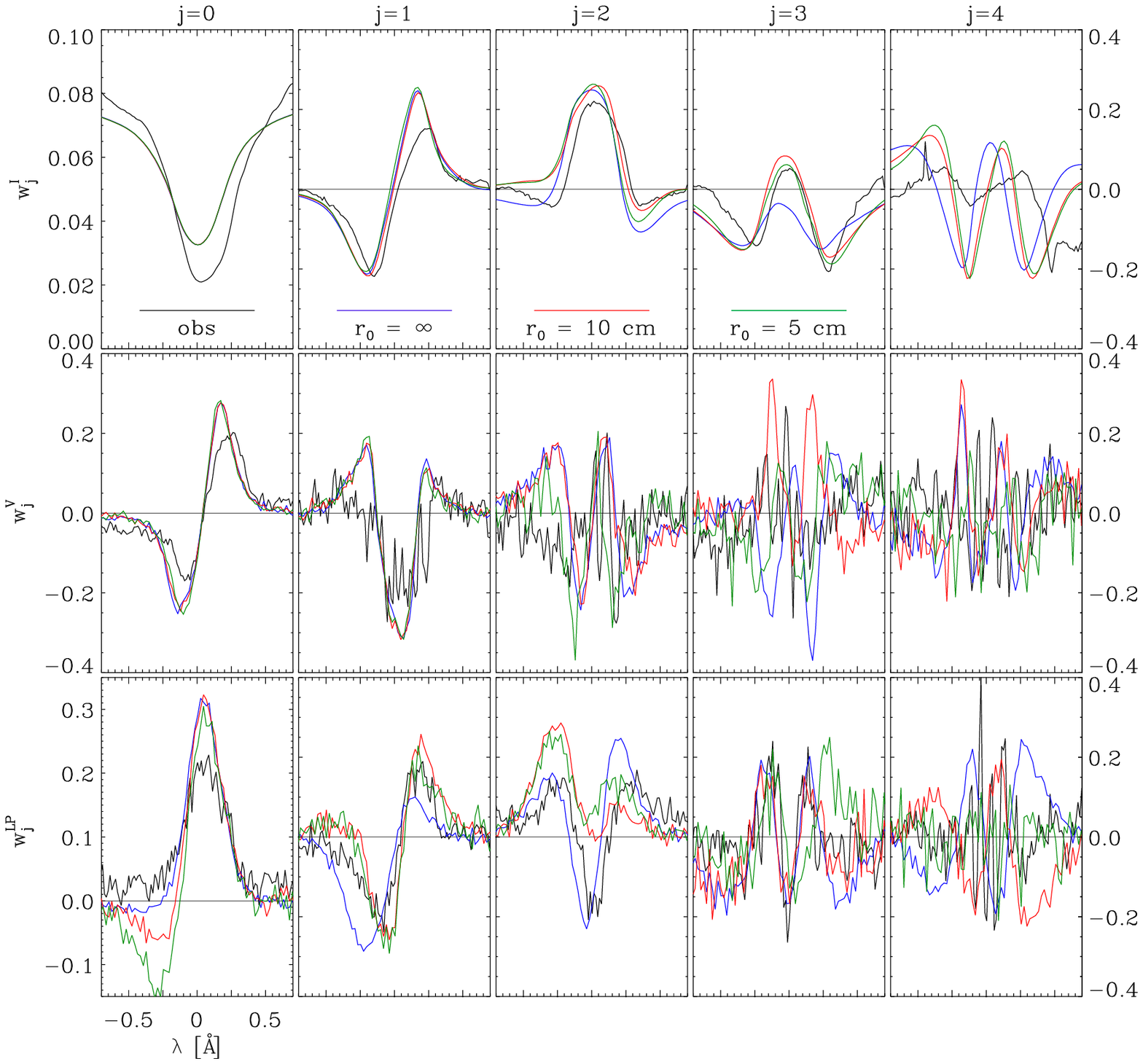}
 \caption{Comparison of the first eigenvectors resulting from the PCA analysis for the observed and theoretical Stokes $I$ profiles (top), $V$ profiles (middle), and linear polarisation profiles (bottom). The black curves correspond to observations, the blue curves to the theoretical profiles that are not degraded with a spatial PSF, and the red and green curves represent the theoretical profiles degraded with PSF$_{r_0=10~\mathrm{cm}}$ and PSF$_{r_0=5~\mathrm{cm}}$, respectively.}
 \label{fig_pca}
\end{figure*}

In the right panel of Fig.~\ref{fig_sig_int}, we show the histogram of minimum intensities in the core of the Ca~{\sc ii}~8542\,\AA\ line. The observed intensities show a very small span of values with a maximum likelihood of about $0.83 \times 10^{-5}$ erg cm$^{-2}$ s$^{-1}$ sr$^{-1}$ Hz$^{-1}$, while the distribution of the minima of the synthesised profiles is very broad when they are not spatially degraded. With decreasing values of the $r_0$ parameter, the distribution of the minimum Stokes $I$ intensities becomes narrower. It is most similar to the observed parameter in the case of the synthetic profiles that are degraded with PSF$_{r_0=5~\mathrm{cm}}$. 

In the left plot of Fig.~\ref{fig_sig}, we compare the amplitudes of the observed and synthetic average $|Q/I|$ and $|U/I|$ profiles. The non-degraded synthetic linear polarisation profiles have larger amplitudes on average. Owing to the changing sign of the Stokes $Q/I$ and $U/I$ profiles over the field of view (FOV), there are cancellation effects when the data are degraded by the spatial PSFs, and the amplitude of the average profile decreases. The most degraded synthetic profiles that visually match the observations have significantly lower intensities. The theoretical averaged profile degraded with PSF$_{r_0=5~\mathrm{cm}}$ has a double peak that is not observed. 

As shown in the left panel of Fig.~\ref{fig_macr}, the average theoretical profile degraded with PSF$_{r_0=10~\mathrm{cm}}$ has a shape that is similar to the averaged observed profile when macroturbulent broadening is considered. However, the right panel of this figure shows that the amplitude of this average theoretical profile is 1.6 times smaller than that of the observed profile (2.8 times when macroturbulent broadening is considered).

In the right panel of Fig.~\ref{fig_sig}, we show the histogram of the maximum absolute values of the $Q/I$ and $U/I$ profiles. The maximum values observed along the slit are similar to those in the theoretical profiles except for those degraded with PSF$_{r_0=5~\mathrm{cm}}$. The minimum values match as well, but they are defined by the noise level that is set to match the observations. The main difference is the frequency of the enhanced signals in the observations compared to the theoretical profiles. The distribution of the observed signals is wide, with most common values between 0.07 and 0.12\,\%. On the other hand, independent of the spatial degradation, all the theoretical profiles have a narrow signal distribution, with a peak at about 0.07\,\%, and with a long tail towards stronger signals.

This discrepancy in the amplitude of the linear polarisation signals can be interpreted as a lack of sufficient symmetry-breaking effects in the model atmosphere. The maximum theoretical linear polarisation signals are similar to the maximum observed signals, meaning that the symmetry-breaking effects of the model atmosphere are locally strong enough to become similar to the solar chromosphere (neglecting the necessary broadening mechanism). On average, however, the model atmosphere lacks the complexity of the solar chromosphere. 

An additional important aspect of the problem is that our synthetic line profiles are based on a single snapshot of the 3D~MHD simulation. In contrast, the observed profiles result from an observation integrated over 100~seconds. It follows that our synthetic profiles are typically narrower and more asymmetric than when time integration is accounted for. An additional effect of time averaging is that the polarisation amplitudes of the line are reduced. In the disc-centre geometry, the positive and negative linear and circular polarisation signals are equally likely over a long period of time, which leads to their partial or complete cancellation. For a study of the impact of time averaging made using the so-called 1.5D approximation, see \citet{Carlin:2013} and \citet{Carlin:2017}. An important finding of our measurements is that even though the observed data have been deteriorated by time averaging, the polarisation signals are mostly stronger{\em } than the synthetic signals that are computed using an instantaneous model snapshot.

\subsection{Shapes of the linear polarisation profiles}
\label{line_shape}

To study the similarity in shapes of the observed and theoretical profiles in a more quantitative manner, we computed the eigenvectors of the profiles with a principal component analysis (PCA). As shown by \citet{Skumanich:2002} for the case of the Zeeman effect in the solar photosphere, the first eigenvectors can be explained as the perturbations caused by the velocity and vector magnetic field parameters \citep[see also][]{Ariste:2003}. In the very general case of NLTE lines that are also affected by the symmetry-breaking mechanism and by the Hanle effect, the situation is more complicated and the problem of interpreting the PCA components would deserve a detailed study. We can argue, however, that if the resulting eigenvectors have similar shapes for the observed and the theoretical profiles, then the MHD simulation and the line synthesis take the important mechanism into account. In Fig.~\ref{fig_pca} we show the first five eigenvectors for the observed and theoretical Stokes profiles, where the $Q/I$ and $U/I$ profiles are treated together as both have the same  statistical properties at the disc centre. 

In the case of the Stokes $I$ profiles, the eigenvectors of the observed and synthetic line profiles have similar shapes and complexity up to $j=3$ (see upper row in Fig.~\ref{fig_pca}). Only for $w_4^I$, we have a discrepancy in the wavelength position of the observed and synthetic eigenvectors, while their shape and complexity are similar. We also start to see the noise pattern in the observed $w_4^I$. As we did not estimate the noise level in the Stokes $I$ profiles, the synthetic $I$ profiles are not degraded by noise and it cannot appear in their eigenvectors. For the Ca~{\sc ii}~8542 line, the $w_1^I$ is close to the first derivative of the $w_0^I$ and can be associated with the LOS velocity \citep[see][]{Skumanich:2002}. As the Ca~{\sc ii}~8542 line is significantly broader than the photospheric lines used by \citet{Skumanich:2002} and the selected pixels have weak magnetic fields, the eigenvectors $w_j^I$ for $j \geq 2$ cannot be associated with the magnetic field parameters. 

In the case of the observed Stokes $V$ profiles, only the first two eigenvectors have distinctive shapes that are similar to the eigenvectors of the theoretical profiles (see the middle row of Fig.~\ref{fig_pca}). For $j \geq 2$, the eigenvectors of the observed profiles are significantly influenced by noise and do not correspond to the $w_j^V$ of the theoretical profiles. Although we degraded the synthetic $V$ profiles with the same noise level as was found in the observations, the sample of theoretical profiles is almost 2.5 times larger than the sample of observed profiles, and therefore the noise is less pronounced in the eigenvectors that correspond to the theoretical profiles.

In the case of the Stokes $Q/I$ and $U/I$ profiles, the eigenvectors corresponding to the observations have distinctive shapes up to $j=3$ (see the bottom row of Fig.~\ref{fig_pca}). These four eigenvectors are similar in terms of complexity to those corresponding to the theoretical profiles. However, the exact shapes of the eigenvectors of the theoretical profiles significantly depend on the spatial degradation and often the best match between the observed and synthetic eigenvectors is achieved for non-degraded theoretical profiles, especially in case of the most important first eigenvector. Larger samples of both the observed and synthetic profiles would help us clarify the reasons for this. 

The eigenvectors of the observed and theoretical profiles have similar shapes and complexities up to the $j$ values for which the noise level starts to dominate the eigenvectors. This implies that the physics contained in the MHD simulation and the line synthesis code captures the most relevant processes occurring in the solar chromosphere. The spatial degradation of the theoretical profiles probably does not have exactly the same effect on the shape of the profiles as the neglected temporal averaging. Therefore, it will be worthwhile to repeat this analysis using not only a greater sample of profiles, but also taking into account the temporal averaging, or, alternatively, using observations with exposure times not longer than 10~s, which may be achieved with large-aperture solar telescopes like the Daniel K.\ Inouye Solar Telescope (DKIST) or the European Solar Telescope (EST).

\section{Discussion and conclusions}
\label{conclusions}

We have presented a detailed quantitative comparison of the observed and theoretical Stokes profiles of the Ca~{\sc ii}~8542\,\AA\ chromospheric line. The spectropolarimetric observations of a quiet-Sun disc centre region were obtained at IRSOL using the ZIMPOL instrument. The theoretical profiles were obtained by \citet{Stepan:2016} applying the RT code PORTA \citep{Stepan:2013} and using the 3D model atmosphere described in \citet{Carlsson:2016}.

We can briefly summarize our conclusions as follows:
\begin{enumerate}
\item We have presented spectropolarimetric data of the Ca~{\sc ii}~8542\,\AA\ line in the quiet chromosphere using the ZIMPOL spectropolarimeter at IRSOL, providing information on the spatial variations along the direction of the spectrograph slit.
\item As expected, the line intensity profiles are narrower than those synthesised in the 3D~MHD snapshot we used. In addition, we have found that the theoretical linear polarisation profiles are also narrower than the observed profiles.
\item Even though the observed spectra are affected by time averaging, the observed amplitudes of the line-centre fractional linear polarisation signals are, on average, larger then the synthetic signals that were computed using a single snapshot of the model atmosphere.
\item We have performed a first quantitative comparison in terms of the PCA method of the observed theoretical profiles that we calculated using the 3D~NLTE solver.
\end{enumerate}

In spite of the very limited samples of observed and theoretical Stokes profiles, we were able to identify the most significant discrepancies between these samples and to relate them to a possible lack of realism in the currently existing models of solar chromosphere. In agreement with previous studies, we find that the most obvious difference between the observed and theoretical profiles is the line width. All the theoretical Stokes profiles of the Ca~{\sc ii}~8542~\AA\ line are narrower than the observed lines. This implies an insufficient dynamics in the model atmosphere, but there are also other mechanisms that may explain the discrepancy in the line width. The study of \citet{Carlsson:2015} points out that the 3D model atmosphere is not dense enough and an artificially increased column mass broadens the chromospheric line profiles. The width of the theoretical profiles would increase if we had accounted for the temporal averaging of these profiles \citep[e.g.][]{Carlin:2013}.

The amplitudes of the linear polarisation signals of the observed profiles are on average larger than those of the theoretical profiles. This suggests that the model atmosphere we used is not as structured as the solar chromosphere. However, only few theoretical profiles with amplitudes of the Stokes $Q$ and $U$ profiles are similar to the maximum amplitudes of the observed linear polarisation profiles, that is, the model atmosphere has corrugations  in the isosurfaces of temperature, velocity, and of themagnetic field strength and orientation that are strong enough locally. We note that the amplitudes of the theoretical profiles can be further decreased by a broadening mechanism that is necessary to match the widths of the observed profiles, for instance, micro- and macro-turbulence and (or) temporal averaging. 

The similarity of the shapes of the first eigenvectors of the observed and theoretical profiles suggest that the MHD simulation and the line synthesis take into account all the important physical aspects that occur in the solar chromosphere. A comparison like this of the eigenvectors of the observed and theoretical profiles has not been done before, but this particular part of our analysis would greatly benefit from a larger sample of both observed and theoretical profiles.

In our analysis, we focused on the quantitative analysis of the observed and synthetic spectra. The ultimate goal of the spectropolarimetric observations would indeed be to decipher the thermal and magnetic structure of the atmosphere. Given the complicated physics of formation of the 8542 line in a 3D medium, namely the role of the symmetry-breaking effects, it is currently complicated to infer the magnetic field vector solely from this type of observation in quiet-Sun regions. On the other hand, it can be shown that the linear polarisation signals of the line can be used to constrain the geometry of the magnetic field lines at the line formation height. This analysis will be the subject of our next paper.

We wish to stress that suitable observations of forward-scattering polarisation can be achieved with current instruments with reasonable exposure times and that this analysis is based on a test run of the ZIMPOL instrument to estimate its capabilities. We need significantly improved observations, for example, using ZIMPOL attached to the GREGOR telescope \citep{Schmidt:2012}. For this type of analysis, we need to observe a larger sample of profiles, and additional improvement in the spatial and temporal resolution would be also helpful.      

\begin{acknowledgements}
J.\,J. and J.\,\v{S}. acknowledge financial support by the Grant Agency of the Czech Republic through the grant \mbox{16-16861S} and the project \mbox{RVO:67985815}. J.\,T.\,B. acknowledges financial support by the Spanish Ministry of Economy and Competitiveness through projects \mbox{AYA2014-55078-P} and \mbox{AYA2014-60476-P}, as well as the funding received from the European Research Council (ERC) under the European Union's Horizon 2020 research and innovative programme (Advanced Grant agreement No. 742265). M.\,B. acknowledges the financial support by the Swiss National Science Foundation, grant 200020\_169418. The 3D radiative transfer calculations were carried out with the MareNostrum supercomputer of the Barcelona Supercomputing Centre (National Supercomputing Centre, Barcelona, Spain).
\end{acknowledgements}

\bibliographystyle{aa}
\bibliography{manuscript}

\end{document}